\documentclass[aps, prd, amsmath, floats, floatfix, twocolumn,
superscriptaddress, nofootinbib, showpacs]{revtex4-1}

\usepackage{graphicx}
\usepackage{color}
\usepackage{soul}
\usepackage{url}
\usepackage{bm}         
\usepackage{times}
\usepackage{dcolumn}
\usepackage{bm}
\usepackage{epsf}
\usepackage{amssymb}

\newcommand{\beq}{\begin{equation}}
\newcommand{\eeq}{\end{equation}}
\newcommand{\beqn}{\begin{eqnarray}}
\newcommand{\eeqn}{\end{eqnarray}}

\usepackage{color}

\newcommand{\Caltech}{\affiliation{TAPIR, Walter Burke Institute for Theoretical Physics, MC 350-17,
    California Institute of Technology, Pasadena, California 91125, USA}}

\newcommand{\Cornell}{\affiliation{Center for Radiophysics and Space
    Research, Cornell University, Ithaca, New York, 14853, USA}}
\newcommand{\WSU}{\affiliation{Department of Physics \& Astronomy,
	Washington State University, Pullman, Washington 99164, USA}}

\newcommand{\AEI}{\affiliation{Max-Planck-Institut fur Gravitationsphysik, Albert-Einstein-Institut, D-14476 Golm, Germany}}

\newcommand{\UNH}{\affiliation {Department of Physics \& Astronomy, University of New Hampshire, 9 Library Way, Durham NH 03824, USA}}

\begin{document}

\title{Smooth equations of state for high-accuracy simulations of neutron star binaries}

\author{Francois Foucart}\UNH
\author{Matthew D. Duez}\WSU
\author{Alana Gudinas}\UNH
\author{Francois H\'ebert}\Caltech
\author{Lawrence E. Kidder}\Cornell
\author{Harald P. Pfeiffer}\AEI
\author{Mark A. Scheel}\Caltech

\begin{abstract}
High-accuracy numerical simulations of merging neutron stars play an important role in testing and calibrating the waveform models used by gravitational wave observatories. Obtaining high-accuracy waveforms at a reasonable computational cost, however, remains a significant challenge. One issue is that high-order convergence of the solution requires the use of smooth evolution variables, while many of the equations of state used to model the neutron star matter have discontinuities, typically in the first derivative of the pressure. Spectral formulations of the equation of state have been proposed as a potential solution to this problem. Here, we report on the numerical implementation of spectral equations of state in the Spectral Einstein Code. We show that, in our code, spectral equations of state allow for high-accuracy simulations at a lower computational cost than commonly used `piecewise polytrope' equations state. We also demonstrate that not all spectral equations of state are equally useful: different choices for the low-density part of the equation of state can significantly impact the cost and accuracy of simulations. As a result, simulations of neutron star mergers present us with a trade-off between the cost of simulations and the physical realism of the chosen equation of state.
\end{abstract}

\pacs{04.25.dg, 04.40.Dg, 26.30.Hj, 98.70.-f}

\maketitle

\section{Introduction}
\label{sec:intro}

The equation of state of dense, cold matter in the core of neutron stars remains an important unknown in nuclear physics today. The recent detection by  LIGO and Virgo of gravitational waves powered by a neutron star merger has opened a new way to study this problem~\cite{TheLIGOScientific:2017qsa}, and provided some early constraints on the equation of state of neutrons stars~\cite{GW170817-NSRadius,GW170817-PE}. In the future, other bright merger events~\cite{Read2009b,Hinderer2010}, or the combination of many dimmer detections~\cite{DelPozzo:13}, could provide tighter constraints on the physics of dense matter.

To determine the parameters of merging compact objects, including the equation of state of neutron stars, gravitational wave signals have to be matched against template banks of semi-analytical waveform models. These models are typically based on approximate solutions to Einstein's equations that are accurate at large separation, but break down close to merger. As a result, the late-time behavior of models has to be carefully calibrated using general relativistic simulations. This is particularly important when attempting to extract information about the equation of state of neutron stars: the strongest finite-size effect measurable by current gravitational wave detectors is the change in the phase of the waveform due to tidal deformation of the neutron star, which only becomes significant during the last few orbits before merger~\cite{Flanagan2008}. 

Reliably modeling finite-size effects in numerical simulations remains a difficult challenge today. The longest high-accuracy simulations available today are $10-15$ orbits long with typical phase errors of $\sim 0.5$ rad~\cite{2017PhRvD..96h4060K,2018arXiv180601625D,Foucart:2018lhe,Kiuchi:2019kzt}, while the phase difference between neutron star merger waveforms and waveforms without finite-size effects (i.e. binary black hole waveforms with objects of the same masses and spins) is typically a few radians, after allowing for an arbitrary time and phase shift of the waveforms (see e.g.~\cite{Foucart:2018lhe}). The best-measured finite-size parameter is the tidal deformability
\footnote{More accurately, gravitational wave detectors are sensitive to the effective tidal deformability, a linear combination of the tidal deformabilities of the two merging compact objects~\cite{Flanagan2008}.}
\beq
\Lambda_{\rm NS} = \frac{2}{3} k_2 \left(\frac{R_{\rm NS}c^2}{GM_{\rm NS}}\right)^5,
\eeq
with $R_{\rm NS},M_{\rm NS}$ the radius and mass of the neutron star, $c$ the speed of light, $G$ the gravitational constant, and $k_2$ the Love number, which varies slightly with the choice of equation of state. At the moment, the impact of $\Lambda_{\rm NS}$ on waveforms is sufficiently well captured by simulations to trust single-event analyses at the signal-to-noise ratio of GW170817, but it remains unclear how much further these analyses can be pushed without being affected by potential systematic errors in the models due to errors in the numerical waveforms used for calibration. 

Obtaining large numbers of high-accuracy numerical waveforms at a reasonable computational cost thus remains an important objective in numerical relativity today. The recent development of high-order methods for the evolution of the equations of relativistic hydrodynamics has significantly improved the accuracy of neutron star simulations~\cite{Radice:2013hxh}. However, high-order convergence is only really possible if the solution is sufficiently smooth. This is potentially problematic. Many of the equations of state currently used in the numerical relativity community provide either analytical formulae or tabulated values for the pressure $P$ as a function of the baryon density $\rho$, temperature $T$ and, if evolved, the electron fraction $Y_e$. Nuclear-theory based tabulated equations of state, as well as the commonly used piecewise-polytropic family of equations of state (PP hereafter)~\cite{Read:2008iy}, have discontinuities in the first derivative of $P$. Discontinuities in the spatial derivative of the fluid variables are also present at the surface of the neutron star, although at lower densities and thus with a lesser impact on the evolution of the spacetime metric and on the gravitational wave signal. The only smooth equations of state currently used in simulations are part of the simple $\Gamma$-law family of equations of state. These equations of state allow us to evolve a neutron star of chosen mass, spin, and tidal deformability, and may thus be appropriate for high-accuracy simulations of black hole-neutron star systems before disruption of the neutron star. However, $\Gamma$-law equations of state typically have unrealistic mass-radius relationships for neutron stars away from that chosen initial mass, making them a poor choice for the study of unequal mass neutron star-neutron star binary inspiral, neutron star disruption, or neutron star-neutron star mergers.

The lack of smoothness of the equation of state is particularly problematic for codes that rely on spectral methods for the evolution of Einstein's equations, such as our Spectral Einstein Code (SpEC~\cite{SpECwebsite}), or codes based on Discontinuous Galerkin methods~\cite{Bugner:2015gqa,kidder:16}, but codes using high-order finite difference methods are not immune to this problem. Here, we investigate an alternative family of equations of state first proposed by Lindblom~\cite{Lindblom:2010bb}: spectral equations of state, for which $\Gamma=d{\ln{(P)}}/d{\ln{(\rho)}}$ is expanded on a set of smooth basis functions. We first review the theoretical description of these equations of state, including improvements made here to the smoothness of the pressure function, then present a cost-effective implementation of these equations of state, and finally study the cost and accuracy of neutron star merger simulations when using these spectral equations of state in SpEC. 

\section{Spectral Equations of State}
\label{sec:SpectalEOS}

\subsection{Formalism}

Our main objective is to construct equations of state that (a) reasonably approximate the physical properties of neutron stars measurable through gravitational waves emitted before merger; and (b) make efficient use of computational resources for long, high-accuracy simulations of neutron star mergers. Accordingly, we will ignore detailed microphysics (e.g. neutrinos) and magnetic fields. We also assume that the neutron star matter is in neutrinoless beta-equilibrium. We describe the neutron star matter as an ideal fluid with stress-energy tensor
\beq
T^{\mu\nu} = (\rho + u + P)u^\mu u^\nu + P g^{\mu\nu},
\eeq
with $u$ the internal energy density, $u^\mu$ the 4-velocity of the fluid, and $g^{\mu\nu}$ the spacetime metric.
The evolution equations are the conservation of baryon number 
\beq
\nabla_\mu (\rho u^\mu) = 0
\eeq
and of energy-momentum
\beq
\nabla_\mu T^{\mu\nu} = 0,
\eeq
i.e. 5 equations for 6 independent variables $(\rho,P,u,u_i)$. To close the system of equations, we need an equation of state $P(\rho,u)$ that describes the properties of dense nuclear matter. More precisely, we typically consider $(\rho,T,u_i)$ as independent physical variables, with $T$ the temperature. The equation of state then specifies the two functions $P(\rho,T),u(\rho,T)$. 

Before merger, finite temperature contributions to the pressure and internal energy are typically negligible. Accordingly, we first consider a composition-independent equation of state for cold matter, $(P_{\rm cold}(\rho),u_{\rm cold}(\rho))$, and add a simple ad-hoc thermal component later in this section. 
The cold energy density and pressure have to satisfy the first law of thermodynamics for adiabatic evolutions,
\beq
d\left(\frac{u_{\rm cold}}{\rho}\right) = - P_{\rm cold}(\rho) d\left(\frac{1}{\rho}\right),
\eeq
so that the cold equation of state is entirely determined by choices for $P_{\rm cold}(\rho)$ and for the specific internal energy $\epsilon=u/\rho$ at zero density, $\epsilon_{\rm cold}(0)$. Here, we choose $\epsilon_{\rm cold}(0)=0$.\footnote{In general, a choice of $\epsilon_{\rm cold}(0)$ is equivalent to a choice for the mass of a baryon, $m_b$, defined so that the equation $\nabla_\mu (\rho u^\mu)=\nabla_\mu (m_b n_n u^{\mu})=0$, with $n_b$ the number density of baryon, indeed imposes conservation of baryon number. Differences between the masses per baryon of protons, neutrons and heavy nuclei are accounted for in $\epsilon$ as binding energy. For a composition-dependent equation of state, $\epsilon(0,Y_e)$ is a function of $Y_e$ and we cannot set $\epsilon(0,Y_e)=0$ anymore.}

The main idea behind the spectral representation of the neutron star equation of state used in this manuscript was first presented by Lindblom~\cite{Lindblom:2010bb}, who also later proposed an improvement to these equations of state that makes it easier to guarantee their causality~\cite{Lindblom:2018rfr}, i.e. that the sound speed in the fluid remains smaller than the speed of light. In this work we choose a formalism close to the original spectral equations of state~\cite{Lindblom:2010bb}, which allows for efficient numerical evolutions, and simply check that our equations of state are causal.

We write the equation of state as a function of the variable $x=\ln{\left(\frac{\rho}{\rho_0}\right)}$, for some
reference density $\rho_0$. The adiabatic index $\Gamma(x)$ is defined such that, at $T=0$,
\beq
\frac{d\ln{P}}{dx} = \Gamma(x).
\eeq
The pressure at $T=0$ can then be expressed as a function of $x$ and $P_0 = P(x=0)$:
\beq
P(x) = P_0 \exp{\left(\int_0^x \Gamma(\tilde x) d\tilde x\right)}.
\eeq
The first law of thermodynamics gives us
\beq
\frac{du}{dx} - u = P(x),
\eeq
which has the solution
\beq
u(x) = u_0 e^x + e^x \int_0^x d\xi P(\xi) e^{-\xi}.
\eeq
with $u_0 = u(x=0)$, or
\beq
u(x) = u_0 e^x + P_0 e^x \int_0^x d\xi e^{-\xi} \exp{\left[\int_0^\xi \Gamma(\tilde x) d\tilde x\right]}.
\eeq
Equivalently, we can compute the specific energy density from
\beq
\frac{d\epsilon}{dx} = \frac{P}{\rho_0} e^{-x},
\eeq
which gives us
\beq
\epsilon(x)   = \epsilon_0  + \frac{P_0}{\rho_0} \int_0^x d\xi \exp{\left[\int_0^\xi (\Gamma(\tilde x)-1) d\tilde x\right]}.
\eeq
To fully define the equation of state at zero temperature, we need
\begin{enumerate}
\item A choice of reference density $\rho_0$, where we fix the pressure $P_0$ and specific internal energy $\epsilon_0$,
\item A choice for the function $\Gamma(x)$ when $\rho>\rho_0$. We use $\Gamma(x)=\sum_{n=0}^N \gamma_n x^n$, which allows us to compute analytically
the inner integral in our formula for $\epsilon$,
\item A choice of equation of state for $\rho<\rho_0$. We use the polytropic equation of state $P=\kappa_0 \rho^{\Gamma_0}$, $\epsilon=\frac{P}{\rho(\Gamma_0-1)}$. 
\end{enumerate}
The parameters $(\rho_0,P_0,\epsilon_0,\kappa_0,\Gamma_0)$ are not independent. In practice, we generally consider $\Gamma_0$, $\rho_0$, and $P_0$ as our free parameters, together with the $\gamma_i$'s. This fixes the cold part of the equation of state. An important difference between the equations of state used in this work and those of Lindblom~\cite{Lindblom:2010bb} is that we additionally require $\gamma_0=\Gamma_0$ and $\gamma_1=0$. With this choice, discontinuities only appear in the third derivative of the pressure, instead of in its first derivative. While this is not necessary to obtain a well-defined equation of state, or to provide good fits to nuclear physic models, we find that it is a crucial component in order to get high-accuracy evolution of binary neutron star systems at a low computational cost (at least when using the SpEC code). For this first study of spectral equations of state, we also choose $N=3$, the smallest number of free coefficients that we found allows us to generate a reasonably wide range of equations of state. In the end, our spectral equations of state thus have 5 free parameters : $\Gamma_0,\rho_0,P_0,\gamma_2,\gamma_3$. We investigate two values of $\Gamma_0$: $\Gamma_0=2$, which leads to better numerical behavior, and $\Gamma_0=1.35692$, which is a more accurate representation of the low-density behavior of cold neutron star matter. We note that for single polytropes, the choice $\Gamma=2$ leads to higher accuracy evolutions than other choices for $\Gamma$. The reason for this behavior is not fully understood, though it may be related to the fact that the density goes linearly to zero at the surface of the neutron star for $\Gamma=2$ equations of state.

The temperature dependence of the equation of state is approximated by the $\Gamma$-law
\beqn
P(\rho,T) &=& P(\rho,0) + \rho T\\
\epsilon(\rho,T) &=& \epsilon(\rho,0) + \frac{T}{\Gamma_{\rm th} -1}
\eeqn
for some constant $\Gamma_{\rm th}$ (we choose $\Gamma_{\rm th}=1.75$). Different choices of $\Gamma_{\rm th}$ can lead to large differences in e.g. the amount of matter ejected in a neutron star merger~\cite{hotokezaka:13}. The impact of that choice on the pre-merger evolution of the system is, however, expected to be negligible. 

For recovery of the primitive variables and computations of the characteristic speeds of the system, the following partial derivatives are also useful, and provided here for completeness:
\beqn
\frac{\partial P}{\partial \rho}|_T &=& \frac{\Gamma P(\rho,0)}{\rho} + T\\
\frac{\partial P}{\partial T}|_\rho &=& \rho\\
\frac{\partial \epsilon}{\partial \rho}|_T &=& \frac{P(\rho,0)}{\rho^2}\\
\frac{\partial \epsilon}{\partial T}|_\rho &=&\frac{1}{\Gamma_{\rm th} -1}\\
\kappa &=& \frac{\partial P}{\partial \epsilon}|_\rho = (\Gamma_{\rm th} -1) \rho\\
h c_s^2 &=& \frac{\partial P}{\partial \rho}|_\epsilon + \frac{\kappa P}{\rho^2}\\
c_s^2 &=& \frac{\Gamma P + (\Gamma_{\rm th}-\Gamma) \rho T}{\rho h}
\eeqn
with $h=1+\epsilon + P/\rho$ the specific enthalpy. We note that causality requires
\beq
c_s < 1,
\eeq
a condition that more advanced versions of spectral equations of state can automatically satisfy~\cite{Lindblom:2018rfr}, but that in our formalism we have to verify holds true for our choices of free parameters. Practically, we only consider equations of state that satisfy $c_s<1$ up to the density $\rho_{\rm max}$ at the center of a neutron star of mass $M_{\rm max}$, with $M_{\rm max}$ the maximum mass of an isolated, non-rotating neutron star.

\subsection{Numerical Implementation}

The main cost associated with the use of a spectral equation of state is the computation of the integrals in our formulae for $P,u,\epsilon$. Our choice of spectral expansion already significantly reduces the cost of these computations, as the pressure can be explicitly written as
\beq
P(x,T) = P_0 \exp{\left(\Gamma_0 x + \gamma_2 \frac{x^3}{3} + \gamma_3 \frac{x^4}{4}\right)} + \rho T
\eeq
for $x>0$ and
\beq
P(x,T) = P_0 \exp{\left(\Gamma_0 x\right)} + \rho T
\eeq
otherwise. For the specific internal energy, we have
\beq
\epsilon(x,T) = \epsilon_0 + \int_0^x d\xi \frac{P(\xi,0)}{\rho_0} e^{-\xi} + \frac{T}{\Gamma_{\rm th}-1}
\eeq
for $x>0$ and, using $\epsilon(0,0)=0$,
\beq
\epsilon(x,T) = \frac{P(x,0)}{\rho (\Gamma_0-1)} + \frac{T}{\Gamma_{\rm th}-1}
\eeq
for $x<0$. The value of $\epsilon_0$ is set by requiring continuity of $\epsilon$ at $x=0$.
For efficient computation of the integral remaining in our expression for $\epsilon$, we resort to a hybrid between tabulation and Gaussian quadrature.
We precompute to round-off accuracy the values of the specific internal energy at a few points $x_i >0$,
\beq
\epsilon_i = \epsilon(x_i,0) = \epsilon_0 + \int_0^{x_i} d\xi \frac{P(\xi)}{\rho_0} e^{-\xi},
\eeq
with $x_i = i \Delta x$ and then use Gaussian quadrature to evaluate
\beq
\epsilon(x,0) = \epsilon_i + \int_{x_i}^x d\xi \frac{P(\xi)}{\rho_0} e^{-\xi}
\eeq
for $x_i<x<x_{i+1}$. Practically, we find that using $\Delta x \sim 0.5$ and a 6-points stencil for Gaussian quadrature allows us to compute $\epsilon$ to roundoff accuracy while only requiring six computations of $P(x)$ per computation of $\epsilon$. As we use $\rho_0 \sim 10^{-4}$ and $\rho \lesssim 0.005$, the table of $\epsilon_i$ has less than $10$ elements. With these choices, we find that a single time step of evolution is $\sim (10-20)\%$ more expensive with these spectral equations of state than when using PP equations of state.

\subsection{Example equations of state}

In simulations, we are generally not interested in using spectral equations of state with given parameters $(\rho_0,P_0,\Gamma_0,\gamma_2,\gamma_3)$, but instead want to map physical properties of the neutron stars such as their radius or tidal deformability at given masses, or the maximum mass of non-rotating neutron stars. We would also like spectral equations of state to have the ability to mimic nuclear-theory based equations of state for cold stars in neutrinoless beta-equilibrium. Unfortunately, we find that simply fitting the function $\Gamma(x)$ or $P(x)$ to their desired value for nuclear-theory based equations of state leads to spectral equations of state that poorly match the physical properties of the original model, unless a large number of basis functions are used ($N \gtrsim 10$, by which point the equation of state is mathematically smooth, but has sharp features that are not resolved in simulations). 

We find that a more powerful method to produce useful spectral equations of state is to use Marko-Chain Monte-Carlo (MCMC) to explore the 4-dimensional space of equations of state (at fixed $\Gamma_0$ -- an assumption that could be abandoned in the future). When performing that search, we automatically reject equations of state that are not causal, or that have a maximum mass below $1.97M_\odot$. We also check whether an equation of state satisfies nuclear physics bounds on the pressure at $\rho = 10^{14.26},10^{14.48}\,{\rm g/cm^3}$ ($\rho = 0.0003,0.0005$ in units where $G=c=M_\odot=1$), taken from Hebeler et al.~\cite{2013ApJ...773...11H}. These bounds on the pressure are not strictly enforced, but instead used to determine the probability that an equation of state is accepted or rejected by the MCMC chain. Using this technique, we generate $\sim 7500$ equations of state for each of the two chosen values of $\Gamma_0$. Illustrative examples are provided in Fig.~\ref{fig:AllMR} and in the tables in the Appendix, where we have chosen equations of state on a grid of $R^{1.35}_{\rm NS}$ (the radius of a $1.35M_\odot$ neutron star) and $M_{\rm max}$, with spacing $\Delta R^{1.35}_{\rm NS}=0.5\,{\rm km}$, $\Delta M_{\rm max}=0.05M_\odot$. Different constraints could however easily be applied to that dataset, and the full set of equations of state is available upon request.

\begin{figure}
\begin{center}
\includegraphics[width=.49\textwidth]{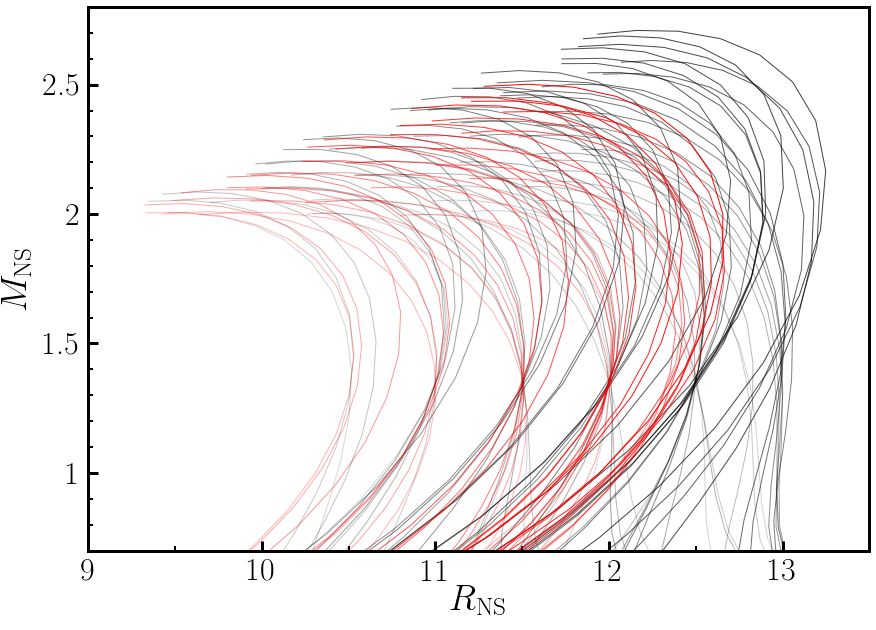}
\caption{Mass-radius relationships for the grid of spectral equations of state with $\Gamma_0=1.35692$ (black) and $\Gamma_0=2$ (red) provided in the Appendix. Both choices of $\Gamma_0$ are compatible with the most likely range of mass and radii, but the lower $\Gamma_0$ choice allows for the construction of stiffer equations of state.}
\label{fig:AllMR}
\end{center}
\end{figure}

In practice, we find that both sets of equations of state allow us to cover the ``most likely'' range of neutron star parameters, as inferred from gravitational wave and electromagnetic observations of GW170817~\cite{GW170817-PE,GW170817-NSRadius,Shibata:2017xdx,2017ApJL2041} and the observation of high-mass neutron stars~\cite{Demorest:2010bx,2013Sci...340..448A}: $M_{\rm max}\in[2,2.3]M_\odot$, $\Lambda \lesssim 800$. The $\Gamma_0=1.35962$ choices however allows for exploration of more extreme equations of state, and a wider range of behavior at fixed $R^{1.35}_{\rm NS}$, $M_{\rm max}$, which may help if attempting to fix a larger number of neutron star properties.

\subsection{Potential uses and limits of spectral equations of state}

Before considering simulations using spectral equations of state, it is useful to review some of the assumptions that went into the construction of our models, and their limitations. Our focus here has been to provide a reasonable physical behavior for the {\it high-density} matter within a neutron star, which is the main driver for the evolution of the spacetime, the inspiral of the binary, and the emission of gravitational waves. This is why we construct equations of state matching nuclear theory at $\rho \sim 10^{14.2-14.5}\,{\rm g/cm^3}$, where some constraints are available, and then allow for any equations of state satisfying causality at higher densities, where we do not have reliable constraints on the pressure. On the other hand, our spectral equations of state are generally poor representation of the physical state of matter at densities $\rho \lesssim 10^{14}\,{\rm g/cm^3}$, especially when choosing $\Gamma_0=2$. This is nearly unescapable: physical equations of state have sharp variations in $\Gamma$ at these densities, and thus we have to make a choice between the smoothness of the equation of state and its physical realism in that region. Spectral equations of state with higher order terms and/or smooth transitions between a low $\Gamma$ at low density and a high $\Gamma$ at high density can be constructed, but if they are to match the physics of neutron stars, they will inevitably lead to rapid variations in $\Gamma$ on length scales that are not resolved in numerical simulations -- thus even if the equation of state is mathematically smooth, it would still lead to slower convergence of the numerical simulations at the resolution that we can practically afford. Finally, our equations of state do not have any composition dependence, and an over-simplified temperature dependence.

From these limitations, we can determine the regimes in which our spectral equations of state will be appropriate to use. These equations of state should be a good description of matter if we are interested in the gravitational wave signal coming from inspiraling binaries, disrupting neutron stars, and possibly merging neutron stars (depending on the impact of thermal effects on the gravitational wave signal during mergers, which has not to our knowledge been clearly quantified so far). They are, however, not appropriate for studies of matter outflows, neutrino-matter interactions, or post-merger accretion disks.

\section{Numerical Simulations}
\label{sec:NR}

\subsection{Initial Data}

\begin{table*}
\begin{center}
\caption{Spectral equations of state used in this work. All quantities are in $G=c=M_\odot=1$ units, except $R^{1.35}_{\rm NS}$ which is given in kilometers (for consistency with tables in the Appendix).}
{
\begin{tabular}{c||c|c|c|c|c|c|c}
Model & $M_{\rm max}$ & $R^{1.35}_{\rm NS}$ & $\Gamma_0$ & $\rho_0$ & $P_0$ & $\gamma_2$ & $\gamma_3$  \\
\hline
SLy$\Gamma$2 & 2.06 & 11.47 & 2 & 1.0118e-4 & 3.3625e-7 & 0.4029 & -0.1008\\
SLy$\Gamma$1.35 & 2.05 & 11.45 & 1.35692 & 8.2235e-5 & 2.5632e-7 & 0.9297 & -0.2523\\
\end{tabular}
\label{tab:eos}
}
\end{center}
\end{table*}

To test the performance of these equations of state, we first construct initial data in the extended conformal thin-sandwich formalism~\cite{York1999,Pfeiffer-York:2005} by solving for the constraints in Einstein's equations, hydrostatic equilibrium, and an irrotational velocity profile for the neutron star's fluid. These equations are solved using our in-house {\sc spells} code~\cite{Pfeiffer2003}, as adapted for binary neutron star systems~\cite{FoucartEtAl:2008,Haas:2016}. For a given system, the first initial data configuration generated by {\sc spells} is a binary in quasi-circular orbit (i.e. with zero radial velocity), which leads to binaries with an eccentricity of a few percents. To reduce eccentricity, we use the iterative procedure of~\cite{Pfeiffer-Brown-etal:2007}. We aim for eccentricities $e\lesssim0.001$. Practically, we find that the orbital parameters leading to low-eccentricity systems do not strongly depend on the neutron star equation of state. When simulating systems that only differ in their chosen equation of state, we thus only need to perform eccentricity reduction for one equation of state, at least when requiring $e\lesssim 0.001$. Here, all simulations are for equal-mass, non spinning neutron stars with $M_{\rm NS}=1.36M_\odot$ and an initial separation of $d_0=53\,{\rm km}$.

\begin{figure}
\begin{center}
\includegraphics[width=.49\textwidth]{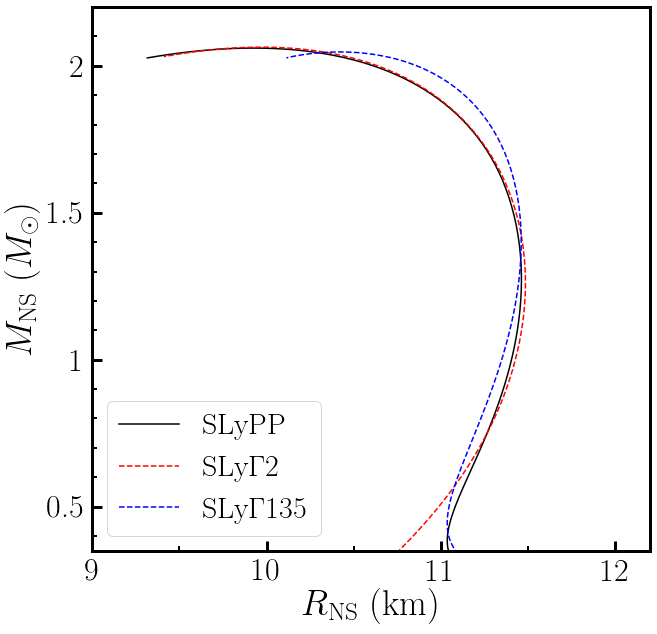}
\caption{Mass-radius relationships for the equations of state from Table~\ref{tab:eos}, and for the piecewise-polytropic equation SLyPP.}
\label{fig:MR}
\end{center}
\end{figure}

In this work, our main objective is to assess the accuracy and computational cost of simulations using different equations of state models. This can usually be done with fairly short simulations. Accordingly, most simulations in this manuscript are only evolved long enough to determine whether the chosen equation of state allows for accurate and cost-effective simulations. We consider three types of equation of state: piecewise polytropic (PP) equations of state, where $\Gamma$ is a piecewise-constant function of density and the first derivative of the pressure is discontinuous; spectral equations of state with $
\Gamma_0=1.35$; and spectral equations of state with $\Gamma_0=2$. For the PP case, we take the SLy equation of state from~\cite{Read:2008iy}, hereafter SLyPP. We then consider spectral equations of state SLy$\Gamma$2 and SLy$\Gamma$1.35 that approximate the SLy equation of state with $\Gamma_0=(2,1.35692)$ respectively (more precisely, we choose spectral equations of state with the same radius at $1.35M_\odot$ and the same maximum mass).  The exact parameters of the spectral equations of state are provided in Table~\ref{tab:eos}, while the corresponding mass-radius relationships are plotted on Fig.~\ref{fig:MR}. We see that the mass-radius relationship of SLy$\Gamma$2 is actually a very good fit to SLy-PP for neutron stars of realistic masses, while it clearly deviates from SLy-PP for lower mass objects. SLy$\Gamma$1.35 is a better fit for low-mass objects, but performs more poorly for higher mass neutron stars. The good match between SLy-PP and SLy$\Gamma$1.35 for low-mass neutron stars is not surprising, as the SLy-PP equation of state also uses a $\Gamma= 1.35692$ polytrope to represents low-density matter.

\begin{figure}
\begin{center}
\includegraphics[width=.49\textwidth]{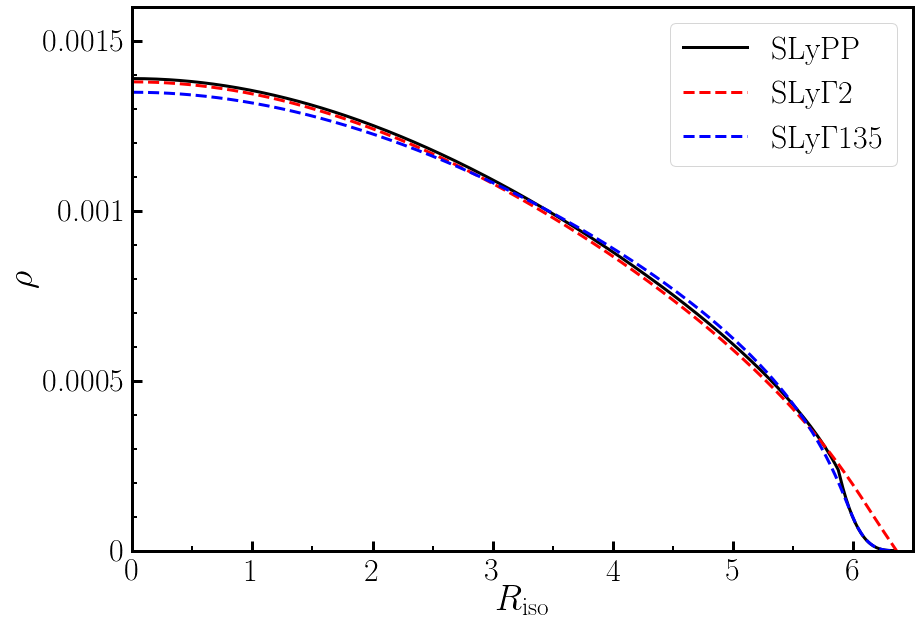}
\caption{Density profile of a $1.35M_\odot$ neutron star for our SLy-like equations of state. Both the density and radius are in $G=c=M_\odot=1$ units, and we plot the density against the radius in isotropic coordinates (i.e. for a conformally flat metric). We note that the isotropic radius differs from the Schwarzschild radius quoted elsewhere in this paper and more commonly used in the literature. We use it here as isotropic coordinates are closer to the coordinates used in our simulations.}
\label{fig:density}
\end{center}
\end{figure}

The different behaviors at low density are also visible on density profiles of neutron stars of fixed masses, e.g. the $1.35M_\odot$ profiles shown on Fig.~\ref{fig:density}. The SLy$\Gamma$2 clearly differs from the other two equations of state at densities below $\rho\sim 10^{14}\,{\rm g/cm^3}$ ($\rho=0.0002$ in geometric units). While all three SLy-like equations of state agree to better than $1\%$ in the radius of a $1.35M_\odot$ neutron star, the different internal structures lead to $\sim 5\%$ changes in the tidal deformability of these neutron stars. Choosing a spectral equation of state that exactly matches a given tidal deformability, instead of a given radius, is however perfectly possible.

\subsection{Numerical Methods}

Binary evolutions are performed with the SpEC code~\cite{SpECwebsite}, using the two-grids method described in more detail in~\cite{Duez:2008rb,Foucart:2013a}. In the two-grid setup, Einstein's equations are evolved on a pseudospectral grid using the Generalized Harmonics formalism~\cite{Lindblom2006}, and adaptive mesh refinement~\cite{Szilagyi:2014fna}. The general relativistic equations of hydrodynamics are evolved using fifth-order shock capturing finite difference methods~\cite{Radice:2012cu}. Both sets of equations are evolved in time using a third-order Runge-Kutta algorithm. The source terms that couple Einstein's equations to the fluid (and vice-versa) are communicated at the end of each full time step, while values of the source terms at intermediate steps of the Runge-Kutta algorithm are obtained by linear extrapolation of their value at the beginning of the current and previous steps. Interpolation from the spectral to finite difference grid is performed by first refining the spectral grid by a factor of $\sim 3$ in each dimension (with a limit of $50^d$ basis functions for a set of basis spanning a $d$-dimensional space), and then using third-order polynomial interpolation from the colocation points of the refined spectral expansion to the finite difference grid (this method is a trade-off between the high cost of a full spectral interpolation, and the low accuracy of a simple polynomial interpolation~\cite{Duez:2008rb}). Interpolation from the finite difference grid to the spectral grid uses fifth-order polynomial interpolation. In the asymptotic regime, this algorithm should provide third-order accurate evolutions, though the interplay between the many numerical methods and orders of convergence involved in each simulation can make the exact order of convergence difficult to assess if the error budget is not dominated by a single source of error.

\subsection{Evolution grids}

In our simulations, the spectral grid is composed of a range of `subdomains' of various shapes that cover the entire computational domain. Neighboring subdomains share a surface boundary, but do not otherwise overlap. Two balls are centered on the neutron stars and surrounded by $10-15$ spherical shells that extend up to about twice the neutron star radius. The wave zone uses 32 spherical shells centered on the center of mass of the binary and covering a range of radii $R=[150,2100]\,{\rm km}$. In between these two regions, filled cylinders cover the axis passing through the center of the compact objects, and hollow cylinders cover the regions farther away from that axis. The coordinate system of the computational grid rotates and contracts so that the center of the neutron stars are approximately fixed in the grid frame. The number of basis functions used for each dimension of a spectral element is chosen adaptively, with the objective to reach a given truncation error for the spectral expansion of the solution~\cite{Szilagyi:2014fna}. For simple equations of state, the truncation criteria is checked every $\sim 50GM/c^3$ at the beginning of the simulation (and more often as the compact objects inspiral)
\footnote{The exact timescale is a multiple of the damping timescale of the control system used to keep the center of the neutron stars fixed on the computational grid~\cite{Duez:2008rb,Hemberger:2012jz}}.
 One of the result of this study is that this does not appear to be sufficient to accurately capture the early evolution of the system for PP or spectral equations of state. Here, we present results where spectral mesh refinement is performed every $\sim 50GM/c^3$, and every $\sim 5GM/c^3$. The target truncation error is a function of the grid spacing $\Delta x$ of the finite difference grid, and scales as $(\Delta x)^5$, so that errors on the spectral grid should decrease at worse as fast as errors on the finite difference grid.

The finite difference grid is also adaptive, in that it only covers regions where the fluid density satisfies $\rho \gtrsim 10^{10}\,{\rm g/cm^3}$. We typically consider three resolutions, with $\Delta x$ decreased by $20\%$ for each increase in resolution. We use setups with initially $\sim (61,76,95,120)$ grid points across the diameter of a neutron star. The lowest three resolutions are typical of what we normally use for single-polytrope equations of state, while the highest resolution is used to test the impact of an increased resolution of the finite difference grid while keeping other parameters constant. As the binary spirals in, the physical grid spacing decreases because the neutron stars are fixed in the grid frame. Every time the physical $\Delta x$ decreases by $20\%$, we construct a new grid with $\Delta x_{\rm new}=1.25 \Delta x_{\rm old}$, thus keeping $\Delta x$ approximately fixed during the evolution (and always smaller than at $t=0$). This expansion of the grid typically happens $2-3$ times per simulation when starting $10-15$ orbits before merger.

\subsection{Cost analysis}

Direct comparison of the cost of simulations at a chosen finite difference resolution and given requested accuracy on the spectral grid can be difficult in SpEC. As the spectral grid used to evolve Einstein's equations changes over time, and tries to reach a set accuracy rather than a set grid spacing, we cannot evaluate the cost-efficiency of an equation of state by simply comparing the accuracy of simulations with fixed initial parameters. The cost of a simulation at fixed target accuracy is not a perfect diagnostics either, as the grid spacing on the finite difference grid used to evolve the equation of hydrodynamics is kept constant when changing equation of state, leading to different errors in each simulation. In practice, we thus have to consider a combination of the cost and accuracy of a simulation in order to draw reliable conclusions. 

We consider two measures of the cost of a simulation. First, we can directly measure the CPU-hrs used by simulations, as long as they were performed on the same machine.  Second, for simulations performed on different machines, we find that the relative size of the time step chosen by our adaptive time stepping algorithm is a good proxy for the relative cost of simulations. This is because the time step is usually set by the adaptively-chosen resolution of the highest-resolution element of the spectral grid, while the cost of a single time step is roughly identical for simulations using neutron stars of similar size and at similar orbital separation. We thus use direct CPU cost measurements for simulations performed on the same machine, and the time step size as its proxy for simulations that are not. The validity of this technique is verified on simulations for which both measures are available below. As for the accuracy, we simply compare the orbital phase at different resolutions. 

As in previous work~\cite{Foucart:2018inp}, we estimate errors with respect to an infinite-resolution simulation by considering the difference between our highest-resolution simulation and an approximation of the infinite-resolution solution obtained by Richardson extrapolation of our finite-resolution results. We perform two Richardson extrapolations : one using the lowest and highest resolution, and one using the middle and highest resolution, and estimate simulation errors from each individually. We then take the worst of these two as our error estimate. Richardson extrapolation is performed by assuming second order convergence of the simulations. From previous experience, this method leads to conservative estimates of the simulation errors. The error in the orbital phase scales with the phase error of the gravitational wave signal, at least when neglecting the error due to extrapolation of the gravitational wave signal to null-infinity. Thus, when determining which simulation is most accurate, the two can be used interchangeably. This is convenient for the tests presented here, as the simulations are not evolved long enough to extrapolate the gravitational wave signal to null infinity. We thus use the orbital motion of the binary to compare the accuracy of different simulations. A good estimate of the phase accuracy of the waveform can be obtained by multiplying the orbital phase error by $2$.

In order to study the cost-efficiency of spectral equations of state, we first compare them to PP equations of state. We then discuss the more technical issue of the impact of domain decomposition and adaptive mesh refinenent settings on the accuracy of the simulations. Finally, we compare different spectral equations of state and show that spectral equations of state with $\Gamma_0=2$ are more cost-efficient than spectral equations of state using the more realistic low-density behavior $\Gamma_0=1.35$.

\subsubsection{Piecewise polytrope vs $\Gamma_0=1.35$ spectral equation of state}

\begin{figure}
\begin{center}
\includegraphics[width=.49\textwidth]{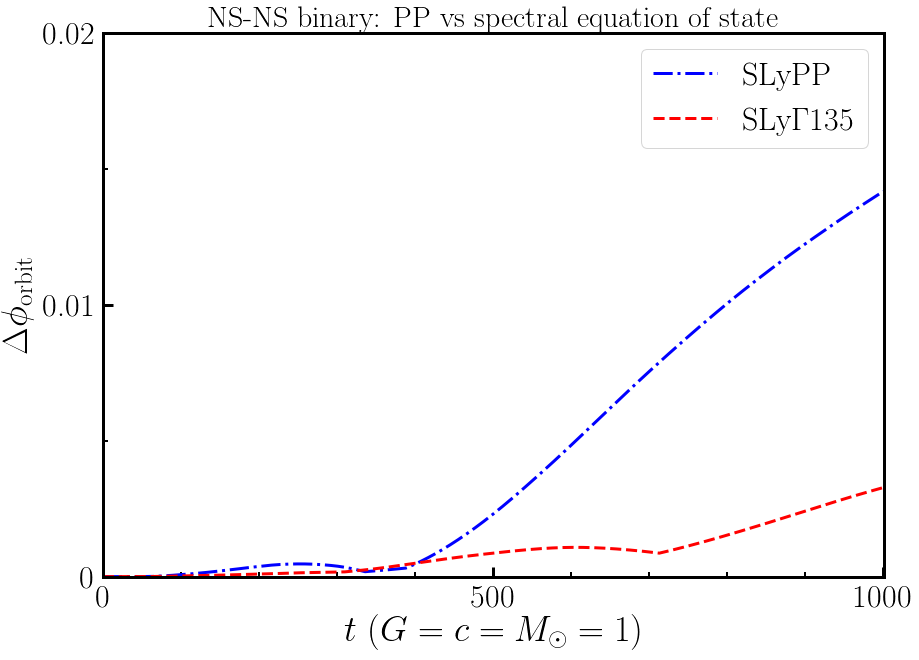}
\caption{Error in the orbital phase of an equal mass NSNS binary evolution as a function of time, using the SLy$\Gamma$135 equation of state and the SLyPP equation of state.}
\label{fig:PhiPP}
\end{center}
\end{figure}

Let us first compare the SLyPP and SLy$\Gamma$1.35 equations of state. Evolving for a little more than an orbit, from $t=0$ to $t=1000$ in units of $GM_\odot/c^3$, costs $(5.9,18.4,50.2)$kCPU-hrs on the Pleiades cluster (Haswell nodes) with the SLyPP equation of state at our 3 highest resolutions, but $(5.7,17.2,32.8)$kCPU-hrs with the SLy$\Gamma$1.35 equation of state. We thus measure comparable costs at low resolution, but a steeper increase in cost at high resolution for the PP equation of state, when the spectral grid tries to capture the sharper features of the neutron star structure. More importantly, the estimated error in the orbital phase at $t=1000$ is $\Delta\phi=0.0045$ rad for SLy$\Gamma$2 and $\Delta\phi=0.014$ rad for SLyPP (see also Fig.~\ref{fig:PhiPP} for the time evolution of the phase error). So the SLy$\Gamma$1.35 evolution is not only slightly cheaper, it is also very significantly more accurate!

We can also use these two sets of simulations to support our argument that the number of time steps at a given resolution is a reasonably proxy for the cost of the simulations. The ratio of the number of time steps in the SLyPP and SLy$\Gamma$1.35 simulations are $(1.23,1.17,1.67)$ at our three resolutions while the ratio of the computational costs are $(1.04,1.07,1.53)$. The slightly larger cost per time step of the spectral equation of state is most likely a consequence of the larger number of operations necessary to compute the internal energy and pressure functions.

The poor performance of the SLy equation of state in SpEC is not a particularly surprising result: PP equations of state were already shown to be very inaccurate in SpEC simulations in~\cite{Foucart:2018inp}. These first results, however, already indicate that spectral equations of state can perform better than PP equations of state at a lower computational cost.

\subsubsection{Domain decomposition and adaptive mesh choices}

So far, the only differences between the simulations previously performed with SpEC for waveform generation~\cite{Foucart:2018inp} and the simulations presented in this work are the use here of a finer resolution on the finite difference grid evolving the neutron star fluid. This was not a particularly inspired change, as it turns out: by varying separately the spectral and finite difference grid, we find that the spectral grid actually dominates the error budget in our simulations with spectral equations of state. This is in contrast with single-polytrope evolutions, for which the spectral evolution of Einstein's equations is typically a subdominant source of error in high-resolution simulations. To improve the accuracy of simulations using spectral equations of state, we make two technical changes to our computational domain. First, we increase the number of spherical shells used around each neutron star from $10$ to $13-15$. When using spectral methods, a lower number of subdomains with higher-order expansion tends to be more efficient is regions where the solution is smooth, while a larger number of subdomains with a lower-order expansion performs better when sharp features are present (here, a discontinuity in the third derivative of the stress-energy tensor).

More importantly, as the surface of the neutron stars is smoothed out in the early evolution and the neutron star itself expands in the coordinates of the computational grid, the relatively small regions of the grid where higher resolution is required to capture sharp features in the structure of the neutron star vary in time. For spectral and PP equations of state, which require neighboring subdomains to have very different number of basis functions, this can be an issue if the adaptive mesh refinement algorithm used to determine the required number of basis functions and, if needed, split into smaller elements subdomains that become too expensive to evolve, is not triggered sufficiently regularly. We find that the rate of trigger of the mesh refinement algorithm in our original simulations (every $\sim 50GM/c^3$) is insufficient, and change to triggering the mesh refinement algorithm every $\sim 5GM/c^3$.

\begin{figure}
\begin{center}
\includegraphics[width=.49\textwidth]{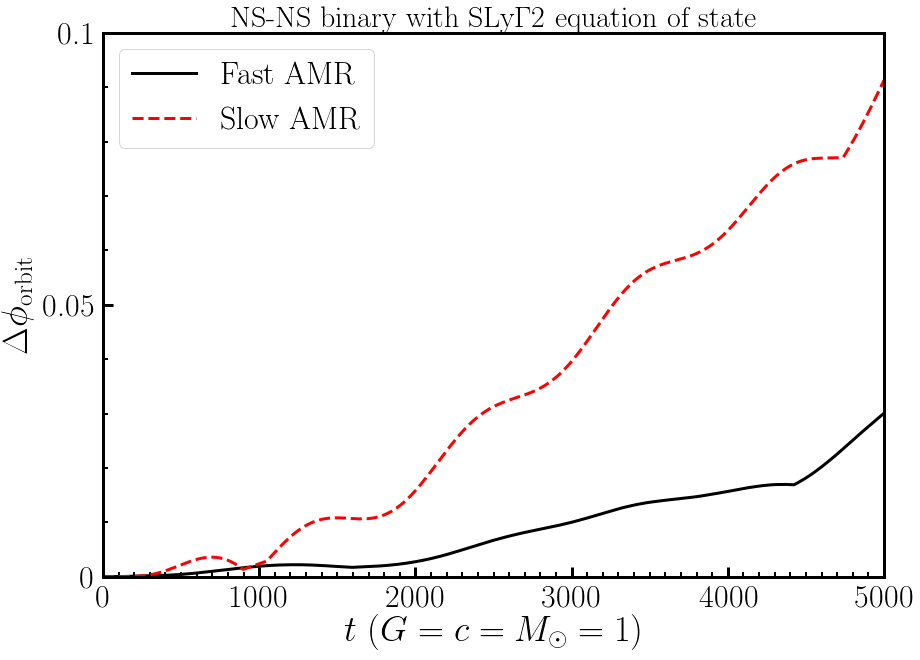}
\caption{Error in the orbital phase of an equal mass NSNS binary evolution as a function of time, using the SLy$\Gamma$2 equation of state and a `fast' or `slow' trigger of the spectral adaptive mesh refinement algorithm. The plot shows $\sim 6.5$ orbits of evolution. Merger would occur after $\sim 10$ orbits, at $t\sim 7000$ in these units.}
\label{fig:PhiAMR}
\end{center}
\end{figure}

To illustrate the resulting accuracy improvement, we consider two sets of simulations with the SLy$\Gamma$2 equation of state, at our three lowest resolutions for the finite difference grid (these will be the standard resolution choices from now on). From low to high resolution, the simulations with frequent mesh refinements have $(1.09,1.13,0.94)$ as many time steps from $t=0$ to $t=1000$ as the simulations with low-frequency mesh refinements, and similar costs as far as can be gathered from simulations performed on different clusters. The orbital phase error is however significantly smaller in the simulation using a more frequent trigger of the mesh refinement algorithm, as shown on Fig.~\ref{fig:PhiAMR}. Some accuracy gains are also observed for the SLy$\Gamma$135 equation of state, although at a higher computational cost, as discussed in the next section.

\subsubsection{$\Gamma_0=1.35$ vs $\Gamma_0=2$ spectral equation of state}

\begin{figure}
\begin{center}
\includegraphics[width=.49\textwidth]{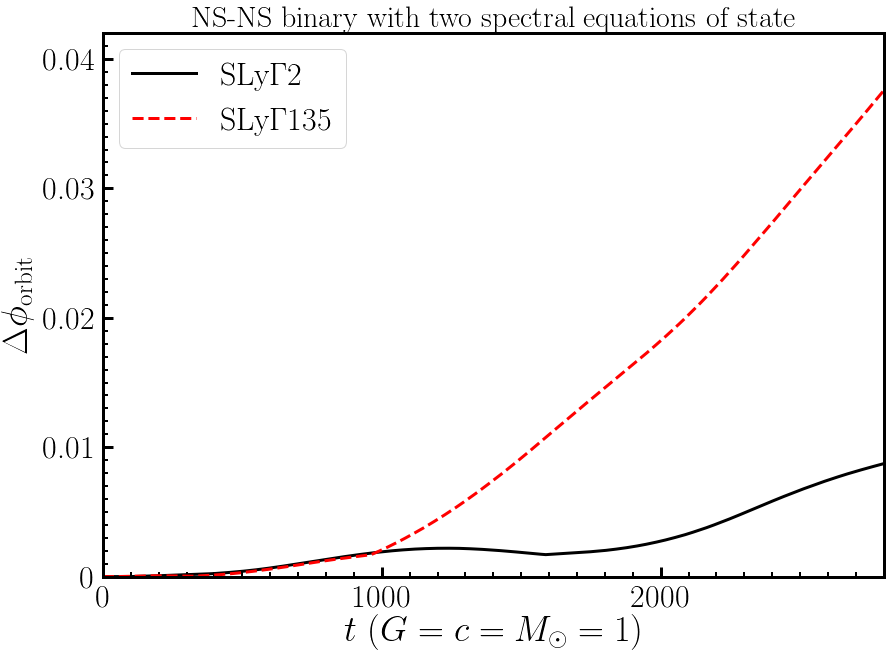}
\caption{Error in the orbital phase of an equal mass NSNS binary evolution as a function of time, using the SLy$\Gamma$2 and SLy$\Gamma$135 equations of state.}
\label{fig:PhiGamma}
\end{center}
\end{figure}

\begin{figure}
\begin{center}
\includegraphics[width=.49\textwidth]{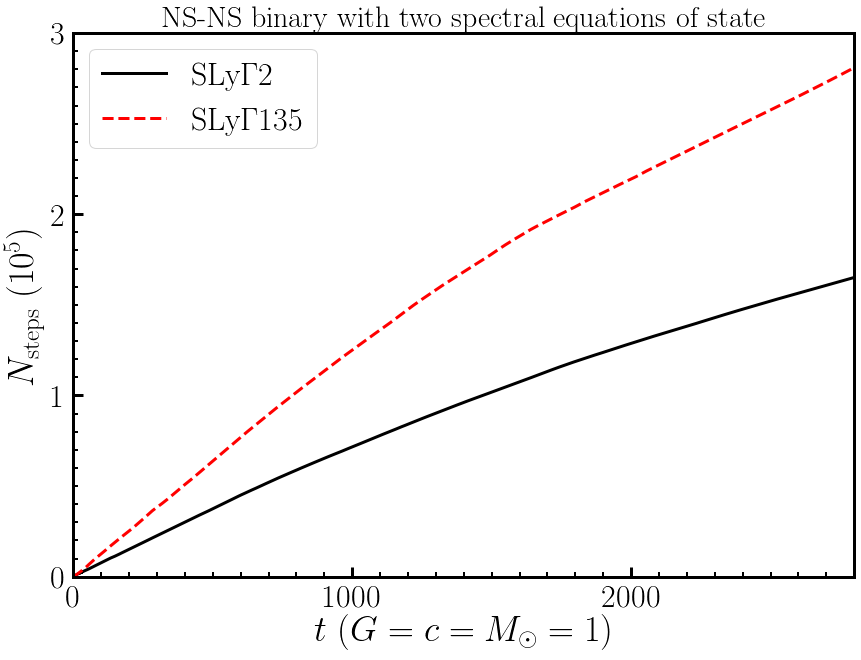}
\caption{Number of time steps taken as a function of time for equal mass NSNS binary evolutions using the SLy$\Gamma$2 and SLy$\Gamma$135 equations of state (high resolution).}
\label{fig:NstepsGamma}
\end{center}
\end{figure}

We can now directly compare the two types of spectral equations of state used in this study. With updated grid choices the accuracy of our  SLy$\Gamma$1.35 simulation is worse than that of the SLy$\Gamma$2 simulation (see Fig.~\ref{fig:PhiGamma}), but still better than the higher resolution PP simulation shown in Fig.~\ref{fig:PhiPP}. The cost of these simulations is also quite different. In Fig.~\ref{fig:NstepsGamma}, we show the number of time steps used by our evolution code for each of these two equations of state, at our highest resolution. We see that the SLy$\Gamma$1.35 requires about twice as many time steps. We emphasize again that this is mostly due to the different grid resolutions chosen by the mesh refinement algorithm to reach that same target accuracy. While most of the spectral grid is nearly identical for these two equations of state, the one shell containing the low-density regions of the neutron star for SLy$\Gamma$2 had to be split into two shells for SLy$\Gamma$1.35, each with about the same number of radial basis functions as the single shell used for SLy$\Gamma$2. As an illustrative example, at $t=800$ and at our highest resolution the total number of points on the spectral grid used by SLy$\Gamma$2 is $9\%$ higher than for SLy$\Gamma$1.35 ($459$k vs $422$k), but the smallest radial spacing in the shells close to the neutron star surface are $\Delta x\sim 0.01$ (SLy$\Gamma$2) and $\Delta x\sim 0.004$ (SLy$\Gamma$1.35), leading to about twice as many steps for SLy$\Gamma$1.35, and only small differences in the cost of a step between the two simulations.

It is of course possible that different choices of grid structure and/or mesh refinement algorithms could lead to smaller phase errors for the SLy$\Gamma1.35$ simulation as well, but this comparison clearly shows that when it comes to reaching a target accuracy at a given computational cost, the SLy$\Gamma2$ equation of state is superior to the SLy$\Gamma1.35$ equation of state.

\subsubsection{Overall cost of the simulations}

We conclude this discussion by looking at the overall cost of the simulations. The high-resolution simulation using slow mesh refinement and the SLy$\Gamma$2 was completed on the Bluewaters cluster, and cost $200$k CPU-hrs up to merger (defined as the first peak of the maximum density after collision of the neutron stars, here $t\sim 7000$ in geometrical units, or $t/M \sim 2600$ with $M$ the total mass of the system). From the scaling discussed in the previous sections, a similar cost is expected for the SLy$\Gamma$135 and SLyPP equations of state with slow mesh refinement, and for the SLy$\Gamma$2 equation of state with fast refinement (the most accurate simulations in this work). The SLy$\Gamma$135 simulation with fast mesh refinement is about twice as expensive, and less accurate.

\section{Conclusions}

We present a first implementation in a general relativistic hydrodynamics code of spectral equations of state meant to capture the high-density behavior of the cold, dense matter in the core of neutron stars. These equations of state are largely inspired by the work of Lindblom~\cite{Lindblom:2010bb,Lindblom:2018rfr}, but modified so that the fluid pressure and internal energy only have discontinuities in their third derivative, including at the matching point between the `low density' and `high density' parts of the equation of state, where the original equations of state had discontinuities in the first derivative of the fluid variables.

We find, at least in the SpEC code, that because spectral equations of state are smoother than traditionally used piecewise-polytrope and tabulated equations of state, their use leads to significantly higher accuracy at a fixed compuational cost. The error in the orbital phase of the binary can easily be improved by factors of a few when using smoother equations of state. This makes spectral equations of state a promising tool for the generation of high-accuracy waveforms that can be used to test existing and future semi-analytical models used for parameter estimation by gravitational wave observatories. The production of such waveforms is in progress.

We also note that while our spectral equations of state can reasonably well capture the potential properties of high-density matter, the most computationally efficient of these equations of state requires the use of an unphysical equation of state for low-density matter. Additionally, all of our spectral equations of state are composition-independent and use an overly simplistic model for the temperature dependence of the fluid variables. While this is not expected to be a major issue for waveform generation and model testing, it indicates that uses of these spectral equations of state for the study of matter outflows and/or post-merger accretion disks would be ill-advised.

\begin{acknowledgements}
F.F. gratefully acknowledges
support from the NSF through grant PHY-1806278,
and from NASA through grant 80NSSC18K0565.
M.D gratefully acknowledges
support from the NSF through grant PHY-1806207.
H.P. gratefully acknowledges support from the
NSERC Canada. L.K. acknowledges support from NSF grant
PHY-1606654 and PHY-1912081. F.H. and M.S. acknowledge support from NSF Grants
PHY-170212 and PHY-1708213. F.H., L.K. and M.S. also thank
the Sherman Fairchild Foundation for their support.
Resources supporting this work were provided by the NASA High-End Computing (HEC) Program 
through the NASA Advanced Supercomputing (NAS) Division at Ames Research Center (allocation HEC-SMD-17-1217).
 This research is also part of the
Blue Waters sustained-petascale computing project, which is
supported by the National Science Foundation (awards OCI-
0725070 and ACI-1238993) and the state of Illinois. Blue
Waters is a joint effort of the University of Illinois at Urbana-
Champaign and its National Center for Supercomputing Applications.
Computations were performed on Trillian, a Cray XE6m-200 supercomputer 
at UNH supported by the NSF MRI program under grant PHY-1229408.
\end{acknowledgements}
%
\section*{References}
\bibliography{References/References.bib}

\appendix
\section{List of spectral equations of state}

\begin{table*}
\begin{center}
\caption{List of spectral equations of state with $\Gamma_0=1.35602$. Radii are in kilometers and masses in $M_\odot$. 
$R^{1.35}_{\rm NS}$ and $\Lambda^{1.35}_{\rm NS}$ are the radius and tidal deformability of a $1.35M_\odot$ neutron star,
and $M_{\rm max}$ the maximum mass of an isolated, non-rotating neutron star.}
{
\begin{tabular}{c|c|c|c|c|c|c|c}
$M_{\rm max}$ & $R^{1.35}_{\rm NS}$ & $\Lambda^{1.35}_{\rm NS}$ & $\Gamma_0$ & $\rho_0$ & $P_0$ & $\gamma_2$ & $\gamma_3$  \\
\hline
1.99992 & 10.5129 & 223.125 & 1.35692 & 0.000127797 & 5.13697e-07 & 1.1343 & -0.323266 \\
2.00108 & 10.997 & 368.996 & 1.35692 & 6.75875e-05 & 1.01558e-07 & 0.987569 & -0.263106 \\
2.00062 & 11.5061 & 364.55 & 1.35692 & 0.000102246 & 4.47771e-07 & 1.03789 & -0.300687 \\
1.99724 & 11.9992 & 523.094 & 1.35692 & 6.37903e-05 & 1.7712e-07 & 0.829266 & -0.221229 \\
2.01024 & 12.5053 & 598.576 & 1.35692 & 7.60147e-05 & 3.06744e-07 & 0.930349 & -0.266615 \\
2.0113 & 12.8074 & 678.516 & 1.35692 & 8.00706e-05 & 3.51137e-07 & 1.03523 & -0.312069 \\
2.04904 & 10.5119 & 239.991 & 1.35692 & 9.35255e-05 & 2.62455e-07 & 0.867112 & -0.216584 \\
2.04463 & 10.9984 & 306.852 & 1.35692 & 9.16453e-05 & 2.93124e-07 & 0.919127 & -0.242781 \\
2.05325 & 11.4962 & 401.966 & 1.35692 & 8.2235e-05 & 2.60576e-07 & 0.9297 & -0.2523 \\
2.04793 & 12.0011 & 445.803 & 1.35692 & 8.44773e-05 & 3.78344e-07 & 0.837305 & -0.225113 \\
2.05043 & 12.4956 & 566.607 & 1.35692 & 8.04569e-05 & 3.64705e-07 & 0.897316 & -0.252764 \\
2.05197 & 12.793 & 678.82 & 1.35692 & 8.17646e-05 & 3.60399e-07 & 1.05283 & -0.316928 \\
2.07671 & 10.6431 & 256.511 & 1.35692 & 0.000101962 & 3.19391e-07 & 0.954798 & -0.248017 \\
2.10149 & 11.0022 & 348.258 & 1.35692 & 7.65352e-05 & 1.59649e-07 & 0.933746 & -0.242265 \\
2.09785 & 11.5059 & 354.424 & 1.35692 & 0.000102789 & 4.79763e-07 & 0.940025 & -0.254744 \\
2.10468 & 11.9997 & 564.734 & 1.35692 & 6.83308e-05 & 1.71472e-07 & 0.970032 & -0.267077 \\
2.10431 & 12.4979 & 714.23 & 1.35692 & 5.69806e-05 & 1.36597e-07 & 0.882562 & -0.238019 \\
2.12443 & 12.8275 & 749.89 & 1.35692 & 6.45398e-05 & 2.16281e-07 & 0.897072 & -0.248748 \\
2.1544 & 11 & 331.297 & 1.35692 & 9.49139e-05 & 2.67417e-07 & 1.0373 & -0.278442 \\
2.1505 & 11.503 & 401.961 & 1.35692 & 9.16938e-05 & 3.20533e-07 & 0.992598 & -0.269978 \\
2.14958 & 11.9972 & 456.526 & 1.35692 & 9.77005e-05 & 4.62847e-07 & 0.997869 & -0.280387 \\
2.15152 & 12.5032 & 571.235 & 1.35692 & 7.9774e-05 & 3.61383e-07 & 0.854295 & -0.23033 \\
2.16462 & 12.9555 & 738.751 & 1.35692 & 8.09392e-05 & 3.568e-07 & 1.05635 & -0.313273 \\
2.19411 & 11.0029 & 358.125 & 1.35692 & 7.92168e-05 & 1.6161e-07 & 0.964422 & -0.247026 \\
2.1931 & 11.5091 & 379.104 & 1.35692 & 0.000106567 & 4.62662e-07 & 1.04415 & -0.286156 \\
2.19819 & 12.0026 & 563.533 & 1.35692 & 4.29011e-05 & 7.96681e-08 & 0.577 & -0.126379 \\
2.20052 & 12.4932 & 643.385 & 1.35692 & 6.97296e-05 & 2.3882e-07 & 0.879873 & -0.236148 \\
2.20431 & 12.9732 & 748.34 & 1.35692 & 8.12841e-05 & 3.58377e-07 & 1.05897 & -0.312069 \\
2.249 & 11.0349 & 359.804 & 1.35692 & 8.6169e-05 & 1.98184e-07 & 1.00877 & -0.260704 \\
2.25359 & 11.5059 & 484.788 & 1.35692 & 6.98537e-05 & 1.30252e-07 & 0.998599 & -0.261916 \\
2.2509 & 12.0114 & 469.033 & 1.35692 & 9.33137e-05 & 4.23167e-07 & 0.921714 & -0.242409 \\
2.24898 & 12.4967 & 792.381 & 1.35692 & 6.2016e-05 & 1.25e-07 & 1.10654 & -0.313508 \\
2.23062 & 12.9943 & 758.146 & 1.35692 & 8.12459e-05 & 3.57726e-07 & 1.05897 & -0.310597 \\
2.28747 & 11.0988 & 401.316 & 1.35692 & 7.15391e-05 & 1.14959e-07 & 0.965624 & -0.241866 \\
2.29791 & 11.5047 & 450.172 & 1.35692 & 7.97909e-05 & 2.03812e-07 & 0.96555 & -0.248432 \\
2.30103 & 11.9988 & 564.263 & 1.35692 & 7.20651e-05 & 1.94531e-07 & 0.938747 & -0.245866 \\
2.30105 & 12.4943 & 691.419 & 1.35692 & 6.81434e-05 & 2.03217e-07 & 0.931841 & -0.24912 \\
2.29735 & 12.9984 & 763.184 & 1.35692 & 8.33722e-05 & 3.72517e-07 & 1.07531 & -0.312832 \\
2.33977 & 11.5138 & 515.102 & 1.35692 & 6.40598e-05 & 9.46418e-08 & 0.992554 & -0.253647 \\
2.35249 & 12.0001 & 646.505 & 1.35692 & 5.58783e-05 & 8.73988e-08 & 0.917981 & -0.232224 \\
2.35168 & 12.4962 & 669.878 & 1.35692 & 8.07974e-05 & 2.88143e-07 & 1.05415 & -0.293308 \\
2.34341 & 12.9902 & 768.911 & 1.35692 & 8.31747e-05 & 3.6599e-07 & 1.0731 & -0.309067 \\
2.40481 & 11.532 & 516.483 & 1.35692 & 7.06947e-05 & 1.18992e-07 & 1.0488 & -0.270769 \\
2.39962 & 11.9969 & 591.885 & 1.35692 & 6.95097e-05 & 1.66063e-07 & 0.939942 & -0.239345 \\
2.40182 & 12.4946 & 687.768 & 1.35692 & 6.31675e-05 & 1.84424e-07 & 0.800003 & -0.195247 \\
2.39386 & 13.0494 & 797.04 & 1.35692 & 8.31747e-05 & 3.64155e-07 & 1.08523 & -0.310953 \\
2.44245 & 11.7225 & 585.851 & 1.35692 & 6.48213e-05 & 9.64132e-08 & 1.04878 & -0.270058 \\
2.45692 & 12.0054 & 683.872 & 1.35692 & 6.02265e-05 & 8.59145e-08 & 1.06694 & -0.279817 \\
2.44811 & 12.5005 & 788.692 & 1.35692 & 5.54727e-05 & 1.08585e-07 & 0.89475 & -0.226102 \\
2.46926 & 12.9902 & 799.071 & 1.35692 & 7.22187e-05 & 2.78566e-07 & 0.902245 & -0.233671 \\
2.48576 & 11.7384 & 592.149 & 1.35692 & 6.48195e-05 & 9.63982e-08 & 1.04269 & -0.266077 \\
2.48565 & 11.9871 & 635.83 & 1.35692 & 7.22697e-05 & 1.46653e-07 & 1.11097 & -0.296504 \\
2.49157 & 12.5042 & 845.198 & 1.35692 & 5.09863e-05 & 7.76165e-08 & 0.92721 & -0.233954 \\
2.50681 & 12.977 & 807.332 & 1.35692 & 7.02376e-05 & 2.60042e-07 & 0.877541 & -0.221707 \\
2.54449 & 12.0157 & 690.912 & 1.35692 & 5.68381e-05 & 7.78775e-08 & 0.973419 & -0.241326 \\
2.54557 & 12.4991 & 860.826 & 1.35692 & 5.64829e-05 & 8.79244e-08 & 1.05925 & -0.279123 \\
2.54063 & 12.8809 & 914.558 & 1.35692 & 6.70385e-05 & 1.71618e-07 & 1.11153 & -0.307275 \\
2.58104 & 12.247 & 780.43 & 1.35692 & 6.13294e-05 & 9.11615e-08 & 1.12897 & -0.299265 \\
2.59919 & 12.5126 & 853.528 & 1.35692 & 5.13518e-05 & 7.90547e-08 & 0.905104 & -0.221663 \\
2.5867 & 12.8726 & 959.423 & 1.35692 & 6.42785e-05 & 1.41709e-07 & 1.15104 & -0.318447 \\
2.63656 & 12.4931 & 847.079 & 1.35692 & 5.59981e-05 & 9.2234e-08 & 0.973824 & -0.243549 \\
2.64711 & 12.9406 & 897.775 & 1.35692 & 7.51349e-05 & 2.33648e-07 & 1.13687 & -0.310468 \\
2.67731 & 12.4766 & 884.743 & 1.35692 & 5.64307e-05 & 7.86026e-08 & 1.08386 & -0.279418 \\
2.69525 & 12.8047 & 959.371 & 1.35692 & 6.12624e-05 & 1.21284e-07 & 1.0879 & -0.285833 \\
\end{tabular}
\label{tab:G1}
}
\end{center}
\end{table*}

\begin{table*}
\begin{center}
\caption{List of spectral equations of state with $\Gamma_0=2$. Radii are in kilometers and masses in $M_\odot$. 
$R^{1.35}_{\rm NS}$ and $\Lambda^{1.35}_{\rm NS}$ are the radius and tidal deformability of a $1.35M_\odot$ neutron star,
and $M_{\rm max}$ the maximum mass of an isolated, non-rotating neutron star.}
{
\begin{tabular}{c|c|c|c|c|c|c|c}
$M_{\rm max}$ & $R^{1.35}_{\rm NS}$ & $\Lambda^{1.35}_{\rm NS}$ & $\Gamma_0$ & $\rho_0$ & $P_0$ & $\gamma_2$ & $\gamma_3$  \\
\hline
2.00446 & 10.5125 & 270.656 & 2 & 8.44019e-05 & 1.20112e-07 & 0.475296 & -0.117048 \\
2.004 & 10.9971 & 334.83 & 2 & 7.91703e-05 & 1.52777e-07 & 0.362182 & -0.0847998 \\
1.99852 & 11.4996 & 418.194 & 2 & 8.2874e-05 & 2.22334e-07 & 0.332367 & -0.0801622 \\
1.99529 & 11.9954 & 640.422 & 2 & 3.60052e-05 & 2.17202e-08 & 0.384077 & -0.093442 \\
2.03575 & 10.5519 & 278.858 & 2 & 8.44019e-05 & 1.20112e-07 & 0.473397 & -0.114594 \\
2.05219 & 11 & 346.539 & 2 & 8.93902e-05 & 1.84618e-07 & 0.446877 & -0.110602 \\
2.05004 & 11.4997 & 427.374 & 2 & 0.000107327 & 3.7473e-07 & 0.469789 & -0.127727 \\
2.04517 & 11.996 & 573.844 & 2 & 6.38051e-05 & 1.22677e-07 & 0.357761 & -0.0903342 \\
2.08233 & 10.7432 & 320.308 & 2 & 0.000110578 & 2.24536e-07 & 0.688799 & -0.19378 \\
2.10218 & 11.0011 & 372.124 & 2 & 6.95251e-05 & 8.21731e-08 & 0.461285 & -0.111624 \\
2.09661 & 11.4995 & 435.963 & 2 & 0.000106155 & 3.52055e-07 & 0.480968 & -0.127724 \\
2.10125 & 11.9967 & 603.129 & 2 & 7.88588e-05 & 1.73422e-07 & 0.524122 & -0.148422 \\
2.14319 & 11.0149 & 378.639 & 2 & 7.74937e-05 & 1.04731e-07 & 0.506452 & -0.124984 \\
2.14992 & 11.5031 & 439.024 & 2 & 0.000141291 & 6.5561e-07 & 0.679557 & -0.202125 \\
2.15009 & 11.9981 & 611.949 & 2 & 6.68359e-05 & 1.16027e-07 & 0.449104 & -0.116738 \\
2.13715 & 12.3413 & 787.297 & 2 & 3.88971e-05 & 2.67158e-08 & 0.432344 & -0.107599 \\
2.20436 & 11.5003 & 492.161 & 2 & 6.27801e-05 & 7.75045e-08 & 0.428688 & -0.100824 \\
2.1977 & 12.0008 & 612.568 & 2 & 7.07278e-05 & 1.33835e-07 & 0.451382 & -0.115274 \\
2.18944 & 12.333 & 745.055 & 2 & 5.40067e-05 & 7.1353e-08 & 0.435096 & -0.111447 \\
2.2587 & 11.4958 & 508.672 & 2 & 6.11326e-05 & 6.44434e-08 & 0.464064 & -0.10905 \\
2.25153 & 12.0012 & 630.262 & 2 & 5.4771e-05 & 6.96652e-08 & 0.375651 & -0.0856468 \\
2.25194 & 12.3631 & 767.079 & 2 & 4.9984e-05 & 5.84238e-08 & 0.40987 & -0.10002 \\
2.2862 & 11.6233 & 551.269 & 2 & 6.17202e-05 & 6.55774e-08 & 0.497713 & -0.120046 \\
2.30679 & 11.9964 & 659.349 & 2 & 4.62256e-05 & 3.95514e-08 & 0.388374 & -0.0870094 \\
2.31224 & 12.3503 & 820.06 & 2 & 4.47172e-05 & 3.37175e-08 & 0.490798 & -0.121708 \\
2.34184 & 12.0112 & 664.427 & 2 & 6.74008e-05 & 9.61116e-08 & 0.537889 & -0.136814 \\
2.35975 & 12.4042 & 814.533 & 2 & 4.63845e-05 & 4.40014e-08 & 0.419602 & -0.0981381 \\
2.40775 & 12.1364 & 735.933 & 2 & 5.13057e-05 & 4.65826e-08 & 0.473559 & -0.111006 \\
2.39412 & 12.4378 & 867.505 & 2 & 4.47172e-05 & 3.36649e-08 & 0.496894 & -0.121708 \\
2.44922 & 12.1751 & 759.508 & 2 & 4.97369e-05 & 4.17553e-08 & 0.480669 & -0.112152 \\
2.43563 & 12.4069 & 826.24 & 2 & 4.65257e-05 & 4.28149e-08 & 0.421593 & -0.0959386 \\
2.49356 & 12.3798 & 844.235 & 2 & 5.36733e-05 & 5.27922e-08 & 0.531654 & -0.129472 \\
\end{tabular}
\label{tab:G2}
}
\end{center}
\end{table*}

\end{document}